\documentclass[twocolumn,twoside,superscriptaddress,pra,aps,showpacs]{revtex4}
\usepackage{amsmath,mathrsfs,amsbsy,color,graphicx,bm,amsthm,amsfonts}
\usepackage{units}
\usepackage{bbm}
\usepackage{times}

\newcommand{\tr}{{\rm Tr}}

\begin{document}

\title{Second law of thermodynamics with quantum memory}
\author{Li-Hang Ren}
\affiliation{Beijing National Laboratory for Condensed Matter Physics, Institute of
Physics, Chinese Academy of Sciences, Beijing 100190, China}
\affiliation{School of Physical Sciences, University of Chinese Academy of Sciences, Beijing 100190, China}
\author{Heng Fan}
\email{hfan@iphy.ac.cn}
\affiliation{Beijing National Laboratory for Condensed Matter Physics, Institute of
Physics, Chinese Academy of Sciences, Beijing 100190, China}
\affiliation{School of Physical Sciences, University of Chinese Academy of Sciences, Beijing 100190, China}
\affiliation{Collaborative Innovation Center of Quantum Matter, Beijing 100190, China}

\begin{abstract}
We design a heat engine with multi-heat-reservoir,
ancillary system and quantum memory. We then derive an inequality related with the second law
of thermodynamics, and give a new limitation about the work gain from the engine by analyzing the entropy change and
quantum mutual information change during the process.
In addition and remarkably, by combination of two independent engines and with the help of the entropic uncertainty relation
with quantum memory, we find that the total maximum work gained from those two heat engines should be larger than a quantity related with quantum
entanglement between the ancillary state and the quantum memory. This result provides a lower bound for the maximum work extracted,
 in contrast with the upper bound in the conventional second law of thermodynamics. However, the validity of this inequality depends on whether the
maximum work can achieve the upper bound.
\end{abstract}

\maketitle

\section{Introduction}
Maxwell's demon plays an important role in the history of thermodynamics and information theory \cite{demon, demon2}.
It is first proposed by Maxwell that a powerful demon might conduct microscopic operation to break the
second law of thermodynamics.
However, according to Landauer's principle, erasure of information will inevitably be accompanied by energy consumption \cite{landauer}, which saves the second law, see also \cite{Bennett}.
Addtionally, with this demon, one can relax the restrictions imposed by the second law on the energy exchanged between a system and surroundings,
and some new thermodynamic inequalities
are studied, see \cite{Lloyd} and a review \cite{Vedral}.
In conventional thermodynamics, the second law gives
$W_{ext}\leq-\bigtriangleup F$, where $W_{ext}$ is the extractable work from system and
$F=U-TS$ is the free energy during the isothermal process.
Due to the Maxwell's demon, this thermodynamic expression can be extended to a favorable form
with discrete quantum feedback control\cite{discrete,minimal}:
\begin{equation}\label{1}
  W_{ext}\leq-\bigtriangleup F+k_B T I,
\end{equation}
in which $k_B$ is the Boltzmann constant and $I$ is the quantum-classical mutual information
describing the mutual information of a fixed quantum system
and the outcome classical information obtained by a quantum measurement.
This quantum-classical mutual information is an extension of the standard quantum mutual information defined originally between
two subsystems. One may then observe that the maximum work that can be extracted may exceed that in conventional
thermodynamics, however, the marginal part is restricted by term of the quantum-classical mutual information.
This inequality lays an extension of the second law of thermodynamics.

The above statement shows information can be exploited to extract physical work,
which may be called an information heat engine \cite{main}. Szilard first explicitly pointed out the
significance of information in thermodynamics, who proposed the so-called Szilard engine (SZE) to realize Maxwell's demon \cite{sze}.
This SZE involves a single-molecule gas in a box, immersed in a thermal reservoir at temperature T,
and an external demon, see also \cite{Vedral}. The demon inserts a partition into the middle of the box,
then measures on which side the molecule is trapped and performs expansion to extract work $W_{ext}=k_B T \ln2$.
In SZE, the information that the molecule is in the left or right is exploited to extract physical work.
Both theoretical and experimental studies on the information heat engine, additionally the extension for quantum case,
have been performed \cite{discrete, minimal, main, sze, entangle,Winter,Plenio,Oppenheim,Wehner,RenLiHang,LloydChenGang,Ueda,review,PNAS}.
Quantum resources for quantum information processing such as quantum entanglement,
quantum discord or quantum coherence may be converted to extractable work.
It is proved that the work gain may result from the entanglement between subsystems \cite{entangle}
because of deep connections between thermodynamics and the theory of entanglement \cite{Winter,Plenio}.
Also one can devise a heat engine which can be driven by purely quantum information
such as the quantum discord \cite{main}. Recently, experimental investigation is performed to show that quantum discord is necessary in energy transport
in a nanoscale aluminum-sapphire interface \cite{LloydChenGang}.

We know that for a quantum system, a quantum memory can be available and
quantum entanglement or some quantumness of correlations may play a critical
role in quantum information processing \cite{book}. With a quantum memory,
the entropic uncertainty relations  generalizing Heisenberg uncertainty principle \cite{heisenberg}
are studied \cite{uncertainty,LiuShang}.
We may notice that entanglement and quantum discord appear to be resource
for work extraction in thermodynamics. It is desirable to construct a heat engine
where quantum correlations or entanglement appear involving in the process.
In this paper, we design a heat engine which includes the system $S$ contacted
by independent heat baths with possible different temperatures, the ancillary
system $A$ and a quantum memory $B$, see Fig. \ref{setup}. This set up of
heat engine is similar with that in Ref.\cite{discrete},
but with an ancillary system $A$ and a quantum memory $B$.
With the help of the quantum memory, we then can characterize the role of quantum
correlation in the thermodynamic circle. Thus, new second law
of thermodynamic inequality can be obtained.

This paper is organized as follows.
In section II,
we design a physical model to realize the information heat engine
and describe the thermodynamic process of our heat engine in detail.
In section \uppercase\expandafter{\romannumeral3},
we derive the extractable work from this engine and
discuss two important cases: the isothermal process and Carnot-like cycle.
In section \uppercase\expandafter{\romannumeral4},
we consider operating processes of two independent engines with different measurement bases.
Based on entropic uncertainty relation with quantum memory,
lower bound of the maximum extractable work with two measurements will be presented.
In section \uppercase\expandafter{\romannumeral5}, we have the conclusion and discussion.

\section{Set up of the heat engine and its process}
\begin{figure}
\centering
\resizebox{0.43 \textwidth}{!}{%
\includegraphics{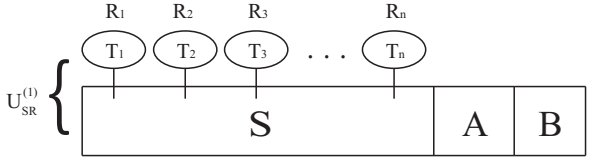}}
\caption{Set up of the heat engine. The system $S$ contacts with independent heat baths
$R_1,R_2,...,R_n$ at temperatures $T_1,T_2,...,T_n$ respectively constituting the multi-heat-reservoir $R$,
the system $S$ has ancillary system $A$, where we can make measurement. Particularly, the set up
includes a quantum memory $B$ which is possibly entangled with ancillary state $A$. At stage (ii), a unitary operation
$U_{SR}^{(1)}$ is performed globally on the system and the multi-heat-reservoir $SR$.}\label{fig:1}
\label{setup}
\end{figure}

The heat engine involves four parts: system $S$, a set of thermal reservoirs $R=\{R_1, \cdots R_n\}$ and
composite quantum system $M$ consisting of ancillary state $A$ and quantum memory $B$, see Fig. \ref{setup}.
The total Hamiltonian is written as
\begin{equation}\label{2}
   H(t)=H_{SR}(t)+H^{int}_{SM_{AB}}(t)+H_{M_{AB}}(t),
\end{equation}
where $H_{SR}(t)=H_S(t)+\sum_{m=1}^n[H_{SR_m}^{int}(t)+H_{R_m}]$,
describing the Hamiltonian of the system, reservoirs and their interaction.
During the operating process, system $S$ can contact $R_1, R_2,\ldots, R_n$
which are at respective temperatures $T_1, T_2,\ldots, T_n$.
By contacting S with $R_1$ at the start and the end of the process,
system $S$ is in thermodynamic equilibrium in the initial and final state.
However, the system may not be in thermodynamic equilibrium between the initial and final state.
For convenience, we note the temperature of S in the initial and final state as $T=T_1$.
At the beginning and last, we assume $H_{SR_m}^{int}(t_i)=H_{SR_m}^{int}(t_f)=0$,
$H_S(t_i)=H_S^{(i)}$ and $H_S(t_f)=H_S^{(f)}$.
The general process of the engine is divided into four stages, which is similar as that
in Ref.\cite{discrete} but with different operations since we have additional quantum
systems $A$ and $B$. This heat engine is also similar as that in Ref.\cite{main} but
with multiple reservoirs in temperatures $T_1,T_2,...,T_n$.

\begin{figure}
\centering
\resizebox{0.43 \textwidth}{!}{%
\includegraphics{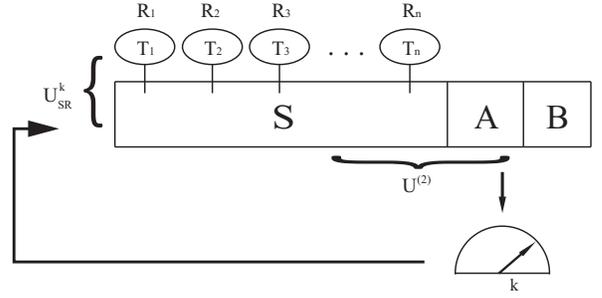}}
\caption{Processing of the heat engine. In the thermodynamic process, at stages (iii) and (iv), a unitary transformation
$U^{(2)}$ is performed on system and ancillary state $SA$, followed by a projective measurement on ancillary
system $A$. Depending on measurement results $k$, a feedback control unitary operation $U^k_{SR}$ is then implemented
on the system and the multi-heat-reservoir $SR$. }\label{fig:1}
\label{process}
\end{figure}

Stage (i). At time $t_i$, the system and reservoirs are in thermodynamic equilibrium respectively, that is,
they are in canonical distribution. The initial state reads
\begin{eqnarray}\label{3}
   \rho^{(i)}&=&\frac{\exp(-\beta H_S^{(i)})}{Z_S^{(i)}}\otimes\frac{\exp(-\beta_1 H_{R_1})}{Z_{R_1}}\otimes\cdots\nonumber\\
    &\otimes&\frac{\exp(-\beta_n H_{R_n})}{Z_{R_n}}\otimes\rho^{(i)}_{AB}   \nonumber\\
    &\equiv& \rho^{(i)}_{SR}\otimes\rho^{(i)}_{AB}   
\end{eqnarray}
where $\beta_n=1/(k_B T_n)$ is related with temperature $T_n$,
and the partition functions are $Z_S^{(i)}=\tr[\exp(-\beta H_S^{(i)})]$ and $Z_{R_n}=\tr[\exp(-\beta_n H_{R_n})]$.

Stage (ii). System $S$ begins to interact with the surrounding to extract work.
In a general way any thermodynamic process between the system and reservoirs
can be expressed by an unitary evolution.
With unitary operation $U^{(1)}=I_{AB}\otimes U^{(1)}_{SR}$ as shown in Fig. \ref{setup}, the initial state becomes
\begin{equation}\label{4}
   \rho^{(1)}=U^{(1)}\rho^{(i)}U^{(1)\dagger},
\end{equation}
in which $I_{AB}$ is the identity operator for $M$ consisting of ancillary state $A$ and memory $B$.
During this process, the composite system $M$ is left unchanged.

In stage (iii), the system $M$ starts to work, which plays the role of Maxwell's demon.
In order to make use of quantum information to extract work, we let the system $S$
interacts with the ancillary system $A$ such that there is quantum information exchange,
then we make measurement on $A$ by positive operator valued measures (POVMs).
Specifically speaking, this measurement process is performed on $A$ with rank-1 projector $\{\Pi_A^k\}$.
So the measurement process is implemented by performing a unitary transformation $U^{(2)}$ on $SA$
followed by a projection measurement $\{\Pi_A^k:|k\rangle\langle k|\}$ on $A$.
The density matrix is given by
\begin{eqnarray}\label{5}
   \rho^{(2)}&=&\sum_k\Pi_A^k U^{(2)}\rho^{(1)}U^{(2)\dagger} \Pi_A^k
  \nonumber \\
   &=&\sum_k p_k |k\rangle_A\langle k|\otimes\rho_{BSR}^{(2)k}
\end{eqnarray}
The measurement outcome $p_k=\tr [\Pi_A^kU^{(2)}\rho^{(1)}U^{(2)\dagger} \Pi_A^k]$ is registered by the memory
and the post measurement state is $\rho_{BSR}^{(2)k}=\tr _A[\Pi_A^kU^{(2)}\rho^{(1)}U^{(2)\dagger}\Pi_A^k/p_k]$.

The stage (iv) is the feedback control. It is performed discretely depending on outcome $p_k$ by
applying corresponding operations on the system $S$ and the multi-heat-reservoir $R$.
Feedback control is also a quantum process, the unitary operator is written as,
\begin{eqnarray}
U^{(3)}=I_B\otimes\sum_k|k\rangle_A\langle k|\otimes U^k_{SR}.
\label{U3}
\end{eqnarray}
The whole process of stage (iii) and the feedback control of stage (iv) are schematically
presented in Fig. \ref{process}.
Now, the final state becomes
\begin{equation}\label{6}
   \rho^{(f)}=U^{(3)}\rho^{(2)}U^{(3)\dagger}.
\end{equation}

In the last equilibration process,
the system and heat reservoirs evolve to reach thermodynamic equilibrium
at temperatures $T$ and $T_1,...,T_n$, respectively, similar as that in \cite{discrete,main}. It
can be described by a unitary transformation $U^{(f)}$ on $\rho^{(f)}$.
That is, the final state is $U^{(f)}\rho^{(f)}U^{(f)\dagger}$.
In the following, entropy calculations will be performed.
However due to the fact, $S(U^{(f)}\rho^{(f)}U^{(f)\dagger})=S(\rho^{(f)})$,
for simplicity, we sometimes write $\rho^{(f)}$ as the final state instead of
$U^{(f)}\rho^{(f)}U^{(f)\dagger}$, if there is no confusion.
Although the system and reservoirs are equilibrium states at last,
which is from a macroscopic point of view,
the final state $U^{(f)}\rho^{(f)}U^{(f)\dagger}$ may not necessarily
be in the form of the rigorous canonical distribution. In order
to evaluate the energy of the system, we introduce the standard canonical distributed
state as the reference state,
\begin{eqnarray}\label{7}
   \rho^{(ref)}_{SR}&=&\frac{\exp(-\beta H_S^{(f)})}{Z_S^{(f)}}\otimes\frac{\exp(-\beta_1 H_{R_1})}{Z_{R_1}}\otimes\cdots\nonumber\\
    &\otimes&\frac{\exp(-\beta_n H_{R_n})}{Z_{R_n}},            
\end{eqnarray}
where $Z_S^{(f)}=\tr[\exp(-\beta H_S^{(f)})]$.
The reference state will be used to correspond the final equilibrium state
$U^{(f)}\rho^{(f)}U^{(f)\dagger}$ with the same temperatures $T$ and $T_1,...,T_n$.

\section{Maximum work extracted from the heat engine}
We will proceed to calculate the net work gain from the heat engine by
analyzing the entropy change during the process.
The von Neumann entropy of the state $\rho$ is defined as $S(\rho)=-\tr(\rho\ln\rho)$.
The following discussion will involve techniques from quantum information \cite{book}.

The difference between states $\rho^{(i)}$ and $\rho^{(1)}$ is a unitary transformation,
so the entropy remains invariant.
However, in the third step, measurement is performed.
Thus, we know that projective measurements increase entropy,
one obtains
\begin{equation}\label{8}
  S[\rho^{(i)}]\leq S[\rho^{(2)}].
\end{equation}
According to equalities (\ref{3}) and (\ref{5}), we know that states
$\rho^{(i)}$ is product state,
$S[\rho^{(i)}]=S[\rho^{(i)}_{SR}]+S[\rho^{(i)}_{AB}]$,
also state $\rho^{(2)}$ can be written as the following form,
\begin{eqnarray}
S[\rho^{(2)}]=H(p_k)+\sum_k p_k S[\rho_{BSR}^{(2)k}].
\end{eqnarray}
With the help of the subadditivity of von Neumann entropy, $S[\rho_{BSR}^{(2)k}]\leq S[\rho_{SR}^{(2)k}]+S[\rho_{B}^{(2)k}]$,
and the above facts, the initial inequality (\ref{8}) now takes the form,
\begin{equation}\label{9}
S[\rho^{(i)}_{SR}]-\sum_k p_k S[\rho_{SR}^{(2)k}]\leq H(p_k)+\sum_k p_k S[\rho_{B}^{(2)k}]-S[\rho^{(i)}_{AB}].
\end{equation}

By considering the form of $U^{(3)}$ in Eq. (\ref{U3}) and tracing out the subsystems $AB$, the final state can be extracted as,
\begin{equation}\label{10s}
 \rho^{(f)}_{SR}= \tr_{AB}[\rho^{(f)}]=\sum_k p_k U^k_{SR}\rho_{SR}^{(2)k}U^{k\dagger}_{SR}.
\end{equation}
Considering the concavity of the von Neumann entropy, we have
\begin{equation}\label{11s}
  S[\rho^{(f)}_{SR}]\geq\sum_k p_k S[\rho_{SR}^{(2)k}]
\end{equation}
By means of taking partial trace like equation (\ref{10s}), we can easily obtain
$\rho^{(f)}_A$, $\rho^{(f)}_B $ and $\rho_{AB}^{(2)}$, and the results are listed as,
\begin{equation}\label{12s}
   S[\rho^{(f)}_A]=S[\rho^{(2)}_A],\qquad S[\rho^{(f)}_B]=S[\rho^{(2)}_B],
\end{equation}
\begin{equation}\label{13s}
   S[\rho^{(2)}_{AB}]=H(p_k)+\sum_k p_k S[\rho_{B}^{(2)k}].
\end{equation}
We then substitute equations (\ref{11s}), (\ref{12s}) and (\ref{13s}) into Eq. (\ref{9}). It is simple to show,
\begin{eqnarray}\label{14s}
   S[\rho^{(i)}_{SR}]-S[\rho^{(f)}_{SR}] &\leq &S[\rho^{(i)}_{SR}]-\sum_k p_k S[\rho_{SR}^{(2)k}]   \nonumber\\
   &\leq & H(p_k)+\sum_k p_k S[\rho_{B}^{(2)k}]-S[\rho^{(i)}_{AB}]
   \nonumber \\
   &=&S[\rho_{AB}^{(2)}]-S[\rho^{(i)}_{AB}] \nonumber\\
   &\equiv &\bigtriangleup S_A+\bigtriangleup S_B-\bigtriangleup I .                   
\end{eqnarray}
Explicitly, the entropy change between initial state and the final state for system $S$ and the multi-heat-reservoir $R$ is
written as,
\begin{eqnarray}\label{14}
   S[\rho^{(i)}_{SR}]-S[\rho^{(f)}_{SR}] &\leq &
   \bigtriangleup S_A+\bigtriangleup S_B-\bigtriangleup I,
\end{eqnarray}
where  $\bigtriangleup S_A=S[\rho_A^{(f)}]-S[\rho_A^{(i)}]$ denotes the entropy
change for ancillary system $A$, and $\bigtriangleup S_B$ denotes the
entropy change of the quantum memory $B$. And
$I$ denotes quantum mutual information defined as
$I(X:Y)=S(\rho _X)+S(\rho _Y)-S(\rho _{XY})$,
here $\bigtriangleup I=I(A^{(2)}:B^{(2)})-I(A^{(i)}:B^{(i)})$ represents the change of quantum
mutual information in the process of heat engine for composite system $M$ including both $A$ and $B$.

According to Klein's inequality $S(\rho\|\sigma)=\tr (\rho\ln\rho)-\tr (\rho\ln\sigma)\geq0$,
where $S(\rho\|\sigma)$ is the relative entropy, we have
$\tr [\rho_{SR}^{(f)}\ln \rho_{SR}^{(ref)}]\leq - S[\rho_{SR}^{(f)}]$.
Therefore with the help of relation (\ref{14}),
\begin{equation}\label{15}
  S[\rho^{(i)}_{SR}]+\tr [\rho_{SR}^{(f)}\ln \rho_{SR}^{(ref)}]\leq \bigtriangleup S-\bigtriangleup I,
\end{equation}
with $\bigtriangleup S\equiv\bigtriangleup S_A+\bigtriangleup S_B$.
From Eq.(\ref{3}) and Eq.(\ref{7}), we know that $\rho^{(i)}_{SR}$ and $\rho_{SR}^{(ref)}$  are canonical distribution.
Then substituting their specific expression into inequality (\ref{14}), we obtain
\begin{eqnarray}\label{16}
    &&E_S^{(i)}-E_S^{(f)}+\sum^n_{m=1}\frac{T}{T_m}(E_{R_m}^{(i)}-E_{R_m}^{(f)})
    \nonumber \\
    &\leq &F_S^{(i)}-F_S^{(f)}
    +k_B T [\bigtriangleup S-\bigtriangleup I],
\end{eqnarray}
where $E_S^{(i)}=\tr(\rho^{(i)} H_S^{(i)})$, $E_{R_m}^{(i)}=\tr(\rho^{(i)} H_{R_m})$, $F_S^{(i)}=-k_B T\ln Z_S^{(i)} $,
$E_S^{(f)}=\tr(\rho^{(f)} H_S^{(f)})$, $E_{R_m}^{(f)}=\tr(\rho^{(f)} H_{R_m})$ by
comparing the final state to the reference state, also $F_S^{(f)}=-k_B T\ln Z_S^{(f)} $.
Among them $E_S^{(f)}-E_S^{(i)}\equiv \bigtriangleup U_S$ is the change of the internal energy,
$E_{R_m}^{(i)}-E_{R_m}^{(f)}\equiv Q_m$ is the heat exchange between system and reservoir $R_m$ and
$F_S^{(f)}-F_S^{(i)}\equiv \bigtriangleup F_S$ is the difference in the Helmholtz free energy of system.
Then the above inequality becomes
\begin{equation}\label{17}
    -\bigtriangleup U_S +\sum_{m=1}^n \frac{T}{T_m}Q_m\leq-\bigtriangleup F_S +k_B T[ \bigtriangleup S-\bigtriangleup I].
\end{equation}
This inequality is one of the main results in the present work.

Before interpreting this extension of the second law of thermodynamic inequality in more detail,
two special cases will first be illustrated.

If $n=1$, there is only one reservoir with temperature $T$.
As the work extractable from the engine is defined as
$W_{ext}=-\bigtriangleup U_S+Q=-(E_S^{(f)}-E_S^{(i)})+E_R^{(i)}-E_R^{(f)}$,
the final result  (\ref{17}) reduces to a simple case
\begin{equation}\label{18}
W_{ext}\leq -\bigtriangleup F_S+k_B T \bigtriangleup S-k_B T \bigtriangleup I.
\end{equation}
This inequality is the same as the result of Ref.\cite{main}.
The inequality means that the extracted work can exceed the difference of the free energy which
generalizes the conventional second law of thermodynamics, however,
the marginal part is constrained by the difference of the changes for entropy and the quantum mutual
information $k_B T (\bigtriangleup S-\bigtriangleup I)$.

When $n=2$, the heat engine becomes an analogue Carnot cycle.
We  take two heat baths with respectively temperatures $T_H$ and $T_L$ for consideration: $T_H>T_L$.
After a cycle, we assume that $\bigtriangleup U_S=\bigtriangleup F_S=0$. Because $W_{ext}=Q_H+Q_L$,
we find,
\begin{equation}\label{19}
     W_{ext}\leq\left(1-\frac{T_L}{T_H}\right)Q_H+k_B T_L(\bigtriangleup S-\bigtriangleup I)
\end{equation}
The efficiency of the heat engine is
\begin{eqnarray}
\eta=\frac{W_{ext}}{Q_H}=1-\frac{T_L}{T_H}+\frac{k_B T_L(\bigtriangleup S-\bigtriangleup I)}{Q_H}.
\end{eqnarray}
In contrast with the efficiency of the conventional Carnot cycle, $\eta_{carnot}=1-\frac{T_L}{T_H}$, the
heat engine presented here can exceed the conventional Carnot heat engine, but new restriction
is still imposed as presented in the last term. On the other hand, as discussed in \cite{main},
the engine does not form a cycle because the final memories are not reset to their initial states.
If reset is performed, the effect of last term should vanish.

The general $n$ case heat engine can be considered as simple extension for $n=1,2$, and the inequality
can be considered as the second law of thermodynamics with quantum correlation in quantum information science.
We comment that the mutual information can be divided into two parts as classical correlation and quantum discord
\cite{Discord}, so $\bigtriangleup I=\bigtriangleup J+\bigtriangleup \delta$.
From the Eq.(\ref{5}), we have $\rho_{AB}^{(2)}= \sum_{k}p_k|k\rangle_A\langle k|\otimes \rho_{B}^{(2)k}$
which is a post-measurement density matrix, so discord $\delta(B^{(2)}|A^{(2)})=0$. Thus for example $n=1$, Eq.(15) becomes
\begin{equation}\label{a}
    W_{ext}\leq -\bigtriangleup F_S+k_B TC,
\end{equation}
where we use the notation, $C=\bigtriangleup S-\bigtriangleup J+\delta(B^{(i)}|A^{(i)})$.
It shows that the initial quantum discord can be exploited to acquire physical work, in agreement
with the results in \cite{LloydChenGang}. Similar form can also be obtained
for general $n$ from Eq.(\ref{17}).

\section{Lower bound for work gained with different measurement bases}
The heat engine in this work includes measurement process at the third stage.
It is well-known that Heisenberg has asserted a fundamental limit to the precision of the outcomes
for a pair of incompatible obervables \cite{heisenberg}.
For a quantum system which can be entangled with a quantum memory, there exists
the entanglement-assisted entropic uncertainty relation for two incompatible measurements \cite{uncertainty},
and more generally for multiple measurements \cite{LiuShang}.
Next, we will study the heat engine,
different from the previous results, by considering two measurements at stage (iii).
For convenience, we just consider the single-reservoir case.

We emphasize that we next concentrate on the measurement process. The heat engine will work twice each with a set of measurement operators,
$K\equiv \{\Pi_A^k: |k\rangle\langle k|\}$ and $M\equiv \{\Pi_A^{\alpha _m}: |\alpha _m\rangle\langle \alpha _m|\}$, however, the whole system should be reset before
the second process begins. On the other hand from a different point of view, we can consider two independent whole systems, the
single measurement will be preformed respectively on each system. Next, we generally consider this condition.
We denote the maximal overlap of the two sets of projective operators as, $c={\rm max}_{k,m}|\langle k|\alpha _m\rangle |^2$.
For state $\rho^{(i)}_{AB}$,  we have the entropic uncertainty relation,
\begin{equation}\label{21}
 S(K|B)+S(M|B)\geq \log_2 \frac{1}{c}+S(A^{(i)}|B^{(i)}),
\end{equation}
in which $S(A^{(i)}|B^{(i)})=S(\rho^{(i)}_{AB})-S(\rho^{(i)}_{B})$ is the conditional entropy for initial state
$\rho _{AB}^{(i)}$, we would like to point out that $S(A^{(i)}|B^{(i)})$ can be negative for entangled initial state.
The quantity $S(K|B)$ is the conditional von Neumann entropy
of the post-measurement state after performing the measurement $\{\Pi_A^k\}$ on $A$,
\begin{eqnarray}\label{22}
 S(K|B)&=& S[\sum_k(\Pi_A^k\otimes I_B)\rho^{(i)}_{AB}(\Pi_A^k\otimes I_B)]-S(\rho^{(i)}_{B})
 \nonumber \\
  &=& S[\sum_k p_k|k\rangle\langle k|\otimes\rho_B^{(i)k}] -S(\rho^{(i)}_{B}),
\end{eqnarray}
where $p_k=\tr (\Pi_A^k\rho_{AB}^{(i)}\Pi_A^k)$, $\rho_B^{(i)k}=\tr _A(\Pi_A^k \rho_{AB}^{(i)} \Pi_A^k)/p_k$.
Similarly, $S(M|B)$ can also be defined after measurement $\{\Pi_A^{\alpha _m}\}$ as follows,
$S(M|B) = S[\sum_m q_m|\alpha _m\rangle\langle \alpha _m|\otimes\rho_B^{(i)\alpha _m}] -S(\rho^{(i)}_B)$,
where $q_m=\tr (\Pi_A^{\alpha _m}\rho_{AB}^{(i)}\Pi_A^{\alpha _m})$ and $\rho_B^{(i)\alpha _m}=\tr _A(\Pi_A^{\alpha _m} \rho_{AB}^{(i)} \Pi_A^{\alpha _m})/q_m$.
We then substitute these equalities into relation (\ref{21}),
that is, the outcomes should satisfy the uncertainty relation,
\begin{eqnarray}\label{24s}
   &&S(\sum_k p_k|k\rangle\langle k|\otimes\rho_B^{(i)k}) -S(\rho^{(i)}_{B})
   \nonumber \\
   &&+S(\sum_m q_m|\alpha _m\rangle\langle \alpha _m|\otimes\rho_B^{(i)\alpha _m})-S(\rho^{(i)}_{B})
   \nonumber \\
   &\geq &\log_2\frac{1}{c}+S(\rho^{(i)}_{AB})-S(\rho^{(i)}_{B}).
\end{eqnarray}

By considering the working process of the heat engine, we know that
$\rho_{AB}^{(2)}=\tr _{SR}\rho^{(2)}= \sum_{k}p_k|k\rangle_A\langle k|\otimes \rho_{B}^{(2)k}$.
For partial trace,
$\rho_B^{(i)k}=\tr _A(\Pi_A^k \rho_{AB}^{(i)} \Pi_A^k)$ and
$\rho_B^{(2)k}=\tr _{SR}[\tr _A(\Pi_A^k \rho^{(1)\prime} \Pi_A^k)]
=\tr _A(\Pi_A^k \rho_{AB}^{(i)} \Pi_A^k)$.
Then we have $\rho_B^{(i)k}=\rho_B^{(2)k}$.
Thus the uncertainty relation (\ref{24s}) can be rewritten as,
\begin{equation}\label{25}
  S(\rho^{(2)}_{AB})+S(\rho^{(2)\prime}_{AB})\geq\log_2\frac{1}{c}+S(\rho^{(i)}_{AB})+S(\rho^{(i)}_{B}),
\end{equation}
where $\rho^{(2)\prime}_{AB}$ indicates the corresponding state when we use the
measurement bases of $\{\Pi_A^m\}$.
Here we find that if we use two independent engines, the measurements are different
for their corresponding engines, the combination of two results constitute an
associated bound.

The extractable work has been illustrated in Eq.(\ref{18}).
Note that the relation (\ref{14}) gives
$\bigtriangleup S-\bigtriangleup I=S[\rho_{AB}^{(2)}]-S[\rho^{(i)}_{AB}]$.
Next we denote $W^K_{max}$ as the maximum extractable work of the first engine with projector $K$ represented as $\{\Pi_A^k: |k\rangle\langle k|\}$,
and similarly $W^M_{max}$ is for another engine with the second measurement, then we have,
\begin{eqnarray}\label{26}
  W^K_{\max}&=&-\bigtriangleup F_S+k_B T[S(\rho_{AB}^{(2)})-S(\rho_{AB}^{(i)})],\\
 W^M_{\max}&=&-\bigtriangleup F_S^{\prime}+k_B T[S(\rho_{AB}^{(2)\prime})-S(\rho_{AB}^{(i)})].
\end{eqnarray}
Due to the limitation of the entropic uncertainty relation, $S(\rho^{(2)}_{AB})+S(\rho^{(2)\prime}_{AB})$
satisfies the inequality (\ref{25}).
With the help of the results for $W^K_{\max}+W^M_{\max}$,
the inequality now reads,
\begin{equation}\label{28}
 W^K_{\max}+W^M_{\max}\geq -\bigtriangleup F_S-\bigtriangleup F_S^{\prime} +k_B T[\log_2\frac{1}{c}-S(A^{(i)}|B^{(i)})].
\end{equation}
The term of conditional entropy $S(A^{(i)}|B^{(i)})$ appearing on the right-hand side can be considered
as quantifying the amount of entanglement between $A$ and $B$ for initial state. If it is negative, we know that
$A$ and $B$ are entangled.
We remark that the quantity $S(A^{(i)}|B^{(i)})$ plays a significant role
in quantum information science, see for example \cite{Lloyd,partialWinter,MingLiang,CerfAdam,Coherence} and the references therein.

The inequality (\ref{28}) plays a complete different role in comparing with the inequalities of second law of
thermodynamics (\ref{17}-\ref{19}). Remarkably, the entropic uncertainty which is a representation of the Heisenberg uncertainty
principle implies that there exists lower bound, instead of upper bound as shown in (\ref{17}-\ref{19}),
for the heat engines presented in this work.
That is to say, the {\it maximum} total work gain can be larger than a bound for two measurement bases,
implying that superposition, which is the reason of no-cloning theorem \cite{FanRep}, is related
with extractable work.
The negative $S(A^{(i)}|B^{(i)})$, meaning the existence of entanglement,
will result in higher lower bound and let the work extracted larger
in case of two measurements with each performing on an individual engine. Straightforwardly, those results can be generalized for multiple measurements.
We still assume that we have several independent engines each with different measurements.

On the other hand, the validity of above interpretation about the inequality (\ref{28}) depends on crucial conditions.
The extractable work is upper bounded in the relation (\ref{18}), we assume that the maximum extractable work $W_{\max}$
can saturate this bound. For this saturation, there should be very strong requirements on the choice of measurement basis and the scheme
of feedback control \cite{Kurt,Kurt1,LiuYXRep}. If the saturation cannot be achieved, the inequality (\ref{28}) in general does not hold.

\section{Conclusion and discussion}

We design a specific heat engine with both ancillary state and the quantum memory. The new inequality
related with the second law of thermodynamics is obtained. For simple cases, our results extend the
results in the isothermal process and the Carnot circle. The changes of entropy and the quantum mutual information
lay new limit for the marginal part of work which exceeds the conventional second law of thermodynamics.
In addition, if two measurements are preformed on two independent engines,
we find a new inequality due to the entropic uncertainty relation with the assistance of quantum
memory by combining the results of the two engines.
We also discuss the possibility for the summation of the maximum extractable work of the two engines being
larger than a lower bound. The validity of this inequality depends on whether the maximum extractable work
can saturate its upper bound.

In the study of various extensions of the second law of thermodynamics,
the resources of quantum information such as entanglement and discord can be
shown to play important roles. Experimentally, quantum correlations can be well
measured, however, the work or energy for few particles of microscopic system generally depend on
definition. It is then challenging to check directly the inequalities involving both
quantum correlations and work or energy related with the second law of thermodynamics.
The exploration both theoretically and experimentally are necessary in studying rigorous
testable extensions of the second law in the context of quantum information.

\emph{Acknowledgments:}
This work was supported by MOST of China (Grant Nos. 2016YFA0302104 and
2016YFA0300600), national natural science foundation of China NSFC (Grant Nos. 91536108, 11774406)
and Chinese Academy of Sciences (Grant No. XDB21030300). We thank Zheng-An Wang for helping us in drawing the pictures.



\begin{thebibliography}{99}
\bibitem{demon} J. C. Maxwell, \textit{Theory of Heat}(Appleton, Lonton, 1871).

\bibitem{demon2} H. S. Leff and A. F. Rex, \textit{Maxwell's demons 2}(IOP Publishing, Bristol, 2003).

\bibitem{landauer} R. Landauer, IBM J. Res. Dev. {\bf5}, 183 (1961).

\bibitem{Bennett}C. H. Bennett, Int. J. Theor. Phys. {\bf 21}, 905 (1982).

\bibitem{Lloyd}S. Lloyd, Phys. Rev. A {\bf 56}, 3374 (1997).

\bibitem{Vedral}K. Maruyama, F. Nori, and V. Vedral, Rev. Mod. Phys. 81, 1 (2009).

\bibitem{discrete} T. Sagawa and M. Ueda, Phys. Rev. Lett. {\bf100}, 080403 (2008).

\bibitem{minimal} T. Sagawa and M. Ueda, Phys. Rev. Lett. {\bf102}, 250602 (2009).

\bibitem{main} J. J. Park, K.-H. Kim, T. Sagawa and S. W. Kim, Phys. Rev. Lett. {\bf111}, 230402 (2013).

\bibitem{sze} L. Szilard, Z. Phys. {\bf53}, 840 (1929).

\bibitem{entangle} K. Funo, Y. Watanabe and M. Ueda, Phys. Rev. A {\bf88}, 052319 (2013).

\bibitem{Winter}S. Popescu, A. J. Short, and A. Winter,
Nat. Phys. {\bf 2}, 754 (2006).

\bibitem{Plenio}F. G. S. L. Brandao and M. B. Plenio,
Nat. Phys. {\bf 4}, 873 (2008).


\bibitem{LloydChenGang}S. Lloyd, V. Chiloyan, Y. J. Hu, S. Huberman, Z. W. Liu, and G. Chen,
arXiv:1510.05035.



\bibitem{Ueda}S. Toyabe, T. Sagawa, M. Ueda, E. Muneyuki, and M. Sano,
Nat. Phys. {\bf 6}, 988 (2010).

\bibitem{review} J. M. R. Parrondo, J. M. Horowitz and T. Sagawa,
Nat. Phys. {\bf 11}, 131-139 (2015).

\bibitem{PNAS}F. G. S. L. Brandao, M. Horodecki, H. H. Y. Ng, J. Oppenheim, and S. Wehner,
Proc. Natl. Acad. Sci. U.S.A. {\bf 112}, 3725 (2015).








\bibitem{Oppenheim}J. Oppenheim and S. Wehner,
Science {\bf 330}, 1072 (2010).

\bibitem{Wehner}E. Hanggi and S. Wehner,
Nat. Commun. 4, 1670 (2013).


\bibitem{RenLiHang}L. H. Ren and H. Fan,
Phys. Rev. A {\bf 90}, 052110 (2014).

\bibitem{book} M. A. Nielsen, and I. L. Chuang, \emph{Quantum Computation and Quantum Information}
(Cambridge University Press, 2000).




\bibitem{uncertainty} M. Berta, M. Christandl, R. Colbeck, J. M. Renes and R. Renner, Nature {\bf6}, 659-662 (2010).

\bibitem{LiuShang}S. Liu, L. Z. Mu, and H. Fan,
Phys. Rev. A 91, 042133 (2015).

\bibitem{heisenberg} W. Heisenberg, Z. Phys. {\bf43}, 172 (1927).





\bibitem{Discord}K. Modi, A. Brodutch, H. Cable, T. Paterek, and V. Vedral,
Rev. Mod. Phys. {\bf 84}, 1655 (2012).

\bibitem{partialWinter}M. Horodecki, J. Oppenheim, and A. Winter,
Nature {\bf 436}, 673 (2005).

\bibitem{MingLiang}M. L. Hu and H. Fan,
Phys. Rev. A {\bf 87}, 022314 (2013).

\bibitem{CerfAdam}N. J. Cerf and C. Adami,
Phys. Rev. Lett. {\bf 79}, 5194 (1997).

\bibitem{Coherence}A. Streltsov, E. Chitambar, S. Rana, M. N. Bera, A. Winter, and M. Lewenstein, Phys.
Rev. Lett. {\bf 116}, 240405 (2016).


\bibitem{FanRep}H. Fan, Y. N. Wang, L. Jing, J. D. Yue, H. D. Shi, Y. L. Zhang, and L. Z. Mu,
Phys. Rep. {\bf 544}, 241 (2014).

\bibitem{Kurt}K. Jacobs, Phys. Rev. A {\bf 80}, 012322 (2009).

\bibitem{Kurt1}K. Jacobs, Phys. Rev. A {\bf 67}, 030301 (2003).

\bibitem{LiuYXRep}J. Zhang, Y. X. Liu, R. B. Wu, K. Jacobs, F. Nori, Phys. Rep. {\bf 679}, 1 (2017).
\end{thebibliography}
\end{document}